\documentstyle[epsf]{europhys}

\def\stars{\bigskip\centerline{***}\medskip}

\newif\ifboo \boofalse

\newcommand{\rem}[1]{}
\def \OMIT#1 {{}}
\def \deg#1 {$#1^\circ$}

\begin{document}

\euro{}{}{}{}
\Date{}

\shorttitle{W.-J. ZHU et al: POTENTIALS FOR 6-COORDINATED BORON ETC.}
%
\title
{Effective  potentials for 6-coordinated Boron: structural geometry approach}

\author{W.-J.~Zhu\footnote{Present address: IBM T.J. Watson Research Center  
              P.O. Box 218, Yorktown Heights, NY 10598}
              and C.~L.~Henley}
\institute{Dept. of Physics, Cornell University, Ithaca NY 14853-2501}

\rec{}{}

%
\pacs{
\Pacs{34}{20Cf}
   {Interatomic potentials and forces}
\Pacs{61}{50Lt}
   {Crystal binding; cohesive energy}
\Pacs{71}{20Mq}
   {Elemental semiconductors}
   {Elemental solids}
 \Pacs{33}{15Dj} {Interatomic distances and angles}
      }
         
\maketitle

\begin{abstract}
We have built a database of {\it ab-initio} total energies 
for elemental Boron in over 60 hypothetical crystal structures of varying
coordination $Z$, such that every atom is equivalent.
Fitting to each subset with a particular $Z$, we
extract a classical effective potential, written as a sum 
over coordination shells and dominated by three-atom
(bond angle dependent) terms. 
In the case $Z=6$ (lowest in energy and most relevant), 
the classical potential has a typical error of 0.1 eV/atom, and 
favours the ``inverted-umbrells'' environment seen in real Boron.
\end{abstract}

\section{Introduction}
Boron stands out among the elements 
for the complexity and polymorphous variety of its structures~\cite{donohue}.
\OMIT {Under pressure, this is not quite so; see the recent
discovery of Barium's incommensurate structure.}
The structures are networks built from icosahedrally symmetric 
clusters, and the
typical ``inverted-umbrella'' 
coordination shell is asymmetrical, containing
five neighbors on one side and one on the other.
The stable phase ($\beta$-B) is currently modeled
with 320 atoms/unit cell.
Of the metastable allomorphs~\cite{donohue},
only $\alpha_{12}$ and T$_{50}$ have solved structures~\cite{slack},
the latter of which is stabilized by $4\%$ N atoms
~\cite{kleinman92}. 
\OMIT {amorphous boron is not reliably known.}
The simplest phase $\alpha_{12}$ contains
\OMIT {as confirmed by accurate electronic 
band structure~\cite{kleinman}, 
lattice dynamics studies and force models \cite{shirai}}
3-centered bonds both within and between icosahedra \cite{longuett}.
It was thought that bonding within icosahedra is metallic 
and weaker, while inter-icosahedral bonds are covalent and 
strong~\cite{Beckel,shirai}; however, recent experiments have
questioned this behavior~\cite{Fu99,La99}.

Boron is the only plausible candidate 
for a single-element or covalent quasicrystal, 
which can be speculatively modeled~\cite{Ta95,wey95} 
on either $\alpha_{12}$-B or $\beta$-B.
\OMIT{
Both $\alpha_{12}$-B and $\beta$-B 
are well described as simple approximants of
a hypothetical {\it quasicrystal} phase. }
\OMIT {search for quasicrystal Boron}
This idea inspired the discovery of a new Boron phase~\cite{kimura}, 
as well as the prediction of Boron nanotubes~\cite{boustani}. 

\OMIT {
Our interest is to predict the stable structure -- a challenging
question for any material, since it requires in principle 
comparisons among an infinite set of candidate structures.}
Our motivation is to compare the energies of alternate Boron structures.
Selected crystal structures~\cite{kleinman92,mailhiot}, 
finite icosahedral clusters~\cite{Ta95},
microtube segments, or sheet fragments~\cite{boustani}
were compared by direct {\it ab-initio} calculations.
\OMIT {Much of the existing electronic-structure calculations ([Em90])
focused on electronic excitations. 
Some prior works 
explained bond angles, lengths, and stiffnesses
in a known structure~\cite{kleinman92,shirai}.}
Such computations, however, 
are limited to $O(10^2)$ atoms whereas real Boron has 
more atoms per unit cell, and quasicrystal models require $\sim 10^5$ atoms.
Furthermore Monte Carlo and/or molecular dynamics simulations
are desirable, especially for the liquid~\cite{vast95}.

A tractable approximation is required
of the ab-initio total energy as a function of atomic positions,
such as a classical (many-atom) potential.
This potential must reasonably represent the energy 
not only for the ground-state structure with slight distortions, 
but also for relatively high-energy
local environments,  such as occur in defects or sometimes 
in complex ground states, 
\OMIT {This happens in a ``frustrated'' material, i.e. when it
is mathematically impossible to repeat the locally optimal packing
so as to fill space.}
We know no previous attempts at such a potential. 
(The closest precursor of our work is Lee's
study of a few Boron structures~\cite{SL}, 
comparing atomic-orbital total energies with
a simpler geometrical function of the closed circuits in the 
Boron network.)


We start from a general form of potential into which all Silicon 
potentials~\cite{stillinger,tersoff,chelikowsky,bazant}
could be cast, a sum 
$ E_{total} =\sum_{i}E_i $
over interactions local to each coordination shell:
The site energy $E_i$ of atom $i$ with coordination number
$Z_i$ is broken into terms
depending on one, two, etc. of
its neighbors $j, k, ...$:
\OMIT{I dodge the question if $Z_i$ is an integer here.
In fact, many of these potentials define an effective coordination
of the BOND.}
\begin{equation}
  \label{eq-gen}
    E_i =  \sum_{j}{f_{Z_i}(r_{ij})} +\sum_{ j \neq k}
   {G_{Z_i}(r_{ij},r_{ik},\theta^{(i)}_{jk})} + ...  
   \end{equation}
where $r_{ij}$ is the distance from atom $i$ to $j$, 
and $\theta^{(i)}_{jk}$ the angle formed by
two neighbours $j, k$ and center $i$.
(We shall assume later that
$j, k$ are restricted to ``nearest'' neighbours of site $i$.)

Those Si potentials mostly attempted to capture the role of $Z$ 
and the radial dependences. 
The dependence on bond angles (with fixed coordination
and bond radius) is not so assured, even for Si. 
Three-body terms may be implicit, 
as each bond's effective $Z$ depends on the 
surrounding atoms~\cite{tersoff,chelikowsky}.
Bazant {\it et al}~\cite{bazant} critiqued
the angle dependence of prior potentials, but
they too addressed it indirectly, by assuming various angular
forms and checking which of these gave the most transferable
{\it radial} behaviours.

In this letter, we develop the opposite approach: we isolate the
angular dependence without {\it a priori} assumptions of
analytic form, 
by fitting to a large database in which non-angular 
variables are held fixed. 
We have no hopes that a potential should be
transferable to a structures with 
very different local order; thus our database must be
a large family of structures, among which bond angles
and coordination numbers vary {\it independently}.~\footnote{
We can quantify how well our database samples the
space of possible coordination shells, by use of a 
metric which defines a distance in this space~\cite{PRB}.}
This is not the case for previous databases of
periodic structures.
The reason is, in part, 
that at Boron's typical coordination $Z=6$, 
a much greater variety of coordination shells is plausible
than at ordinary covalent ($Z=3$-$4$) or metallic ($Z\approx 12$) 
coordinations.
\OMIT{
Icosahedral clusters of various scales are
present in all the known atomic arrangements, leading to 
unusual coordination environments and geometric patterns \cite{hoard}, quite
unlike those in the well-studied covalent (e.g. Silicon) systems. }

Consider a database of structures in which
all bond lengths are $R$,  all coordination numbers
are $Z$, and non-pair interactions are limited to nearest neighbors.
Then eq.~(\ref{eq-gen}) reduces to
  \begin{equation}
    \label{eq-Z}
    E_i =  \sum_{j \neq i} f_Z(r_{ij}) +\sum_{ j \neq k}
   {G_Z(R,\theta^{(i)}_{jk})} + h_Z(i)
   \end{equation}
where $G_Z(R,\theta)\equiv G_Z(R,R,\theta)$, 
and $h_Z$ gathers all interactions beyond
the three-body (or rather two-neighbor) terms. 
It is not obvious {\it a priori} what form $h_Z$ should have;
we found empirically (see below) that four-body (three-neighbor) terms are
{\it not} needed, but a $Z$-neighbor term 
$h_Z(i) \equiv c_Z(R) \xi(i)$ 
{\it is} needed. It is proportional to the ``asymmetry'' $\xi(i)$
of coordination shell $i$
(originally introduced to characterize dangling bonds~\cite{chelikowsky}):
  \begin{equation}
  \label{eq-asym}
      \xi(i) =  \frac{| \sum_j {\bf r}_{ij} |}{ (1/Z) \sum_j |{\bf r}_{ij}|}
   \end{equation}
where ${\bf r}_{ij}$  is the vector
from atom $i$ to  its neighbour $j$.
(The denominator is the mean nearest-neighbour distance.)

Now, restrict the database further so 
that every structure is ``uniform''
\OMIT{~\cite{wells},} 
i.e. all sites are crystallographically equivalent and
have the same  local environment; 
\OMIT{\cite{wells}}
We investigate those uniform structures in which all 
nearest neighbour distances are made equal (or nearly so), the better
to separate the $R$-dependence from the angular effects.

Then $E_i$ must be equated with the total energy per atom, 
as found from an LDA calculation.
Now assume that, within the database, only
a discrete set of inter-neighbor angles is possible, $\theta_a$ ($a=1,2,...$);
we realize this approximately by simply dividing the range of 
$\theta$ into several bins.
Then -- if the structures sufficiently outnumber the angular bins --
one obtains $G_Z(R,\theta_a)$ for each angle by a simple linear fit. 
For each value of the scale $R$, the coefficients $\{ G_Z(R,\theta_a)\}$
and $c_Z(R)$ satisfy a set of linear equations:
  \begin{equation}
  \label{eq-linrel}
  E_m(R) - Z f_Z(R) =  \sum_a { N_m(\theta_a) G_Z (R,\theta_a)} + 
                         c_Z(R) {\xi}_m
  \end{equation}
Here $m$ runs over the structures of the same $Z$, 
and $N_m(\theta_a)$ is the number times angle $\theta_a$ 
occurs in the coordination shell of structure $m$.
The whole procedure can be repeated for each $Z$.

\begin{figure}
\epsfxsize=5.1in {\epsfbox{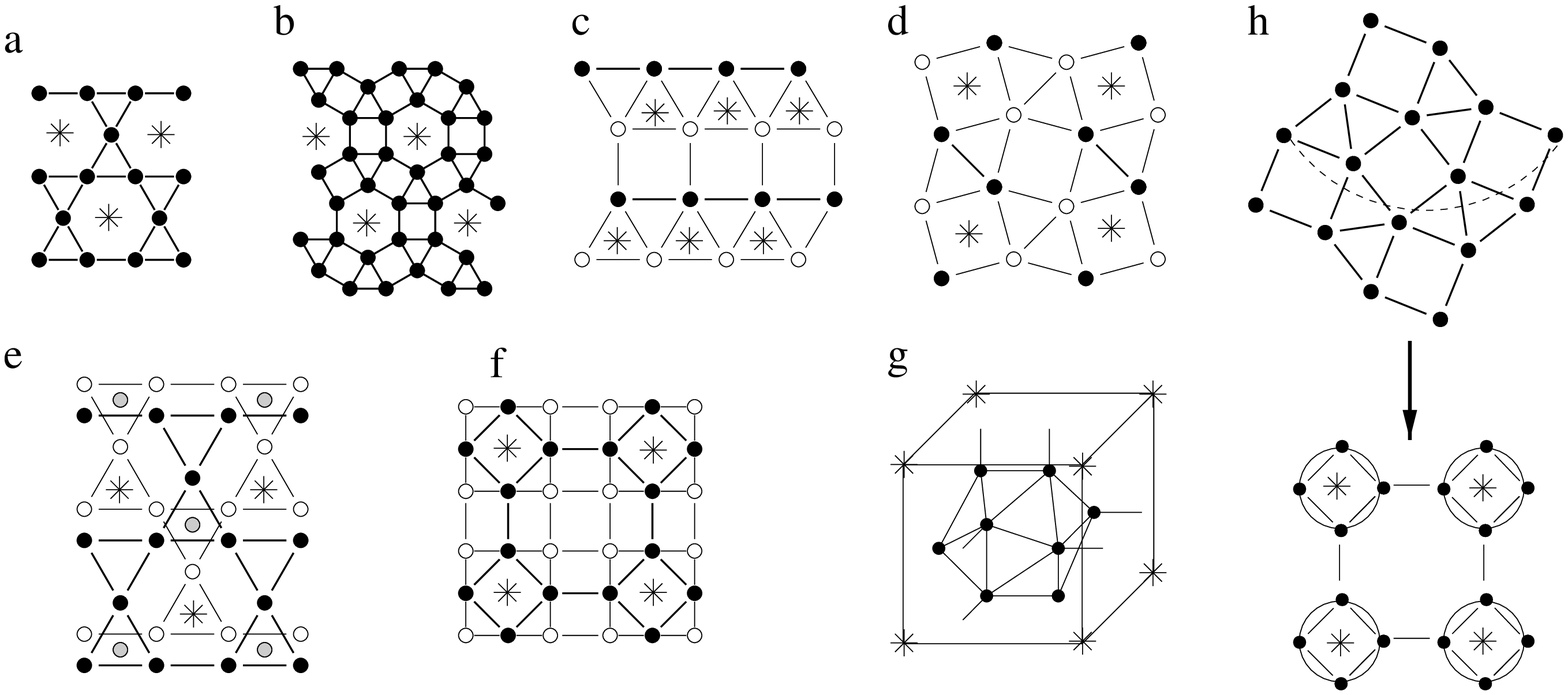}}
\caption{Some less familiar $Z6$ structures. Bravais lattice vertices
are indicated by $\ast$.  For layered structures, atoms
$\bullet$ are nearer to the viewer than $\circ$.  
(a) Kagome, (b) i-dodec, 
(c) $P\mu$, (d) $P\sigma$, 
(e) two layers in T-lattice (the full cubic structure
is an ABC stacking of Kagome layers)
(f) two layers in SC-co (full structure is
$\bullet \bullet \circ$ stacking), 
(g) SC-ico, 
(h) tube-$\sigma$: the $\sigma$-lattice at top is
rolled into tubes (identifying the points connected
by the dashed curve) which are seen end-on at bottom.}
\label{f-struct}
\end{figure}

\section{Structural database}
For our database, we adopted or invented over 60
uniform structures~\cite{PRB}, spanning coordination numbers $Z$ from
3 to 12, and exhibiting a variety of angular patterns for each $Z$; 
\OMIT {being uniform structures,  they are simple to analyze,  yet they
sample very different parts of the configuration
space of coordination shells.}
\OMIT {Certain of our structures are identical to the
B subnetwork in real metalloboride compounds~\cite{SciAm}.}
We shall present details only for 6-coordinated
($Z6$) structures, which are relevant to the real phases:
$80\%$ of the sites in $\beta$-B, and $50\%$ in
$\alpha_{12}$(B), have $Z=6$.
The same procedure was carried out for $Z=4,5$ and $7$.~\cite{PRB}

Our $Z6$ structures, of course, include the simple cubic (SC) 
and the triangular (tri) lattices (parentheses give codes to 
be used henceforth). 
We systematically constructed more $Z6$ structures by
stacking $Z4$ planar lattices, such as the 3.4.6.4
pattern of inter-penetrating dodecagons (i-dodec)
or the 3.6.3.6 lattice (Kagome), see
Figure~\ref{f-struct}(a,b).
We also stacked $Z5$ planar lattices such as 
$3^2.4.3.4$ (called $\sigma$) or the $3^3.4^2$ (called $\mu$), 
puckering alternate layers so that each vertex gained one 
more neighbour from the layer above or below:
this produced the $Z6$ puckered ($P\sigma$) and ($P\mu$) 
structures, respectively as in 
Figure~\ref{f-struct}(c,d).
As far as possible, the inter- and intra-unit bond lengths in
our packings and stackings were made equal.

Figure~\ref{f-struct}(e,f,g)
show inherently three-dimensional networks: 
a simple cubic array of
cuboctahedra (SC-co);  the (T-lattice)
\OMIT{~\cite{okeefe}}
-- also called ``pyrochlore lattice'' --
consisting of a diamond lattice's bond midpoints;
and icosahedra placed on 
a simple cubic lattice (SC-ico).\footnote{
SC-ico was introduced in Ref. \cite{SL}, 
but misidentified  there as ``primitive orthorhombic.''}
We also rolled a triangular lattice
into a single infinite nanotube~\cite{boustani},
with a circumference of 8 edges (tube-tri, not illustrated).
Finally, we rolled the
$Z5$ $3^2.4.3.4$ ($\sigma$) lattice into a tube whose circumference is
the dotted arc in Figure~\ref{f-struct}(h), and packing such
tubes in a square array. This made the ``tube-$\sigma$'' structure, in
which one neighbor of each atom belongs to an adjacent tube.

\begin{table}
\centering
\caption {Local environments of   $Z6$ structures.}
\[
\begin{tabular}{lcccccccccc}
\hline
Structure  &  $N_2$  &  $N_3$  &  \multicolumn{7}{c}{$N(\theta_a)$}& $\xi$\\
  	   & 		& 	&
$\deg{60}$ & $\deg{90}$ & $\deg{108}$ & $\deg{120}$ & $\deg{135}$ &
$\deg{150}$ & $\deg{180}$ & \\
\hline
SC	& 12& 8& 0& 12& 0& 0& 0& 0& 3& 0\\
 tri	& 0&  6& 6& 0&  0& 6& 0& 0& 3& 0\\
SC-co	& 7&  6& 2& 7&  0& 2& 4& 0& 0& 0.586\\
T-lattice & 0& 12& 6& 0&  0& 6& 0& 0& 3& 0\\
i-dodec	& 10& 6& 1& 10& 0& 1& 0& 2& 1& 0.732\\
Kagome	& 8&  4& 2& 8&  0& 2& 0& 0& 3& 0\\
SC-ico	& 1& 10
& 5& 1&  7& 0& 0& 2
& 0& 1.854\\
$P\sigma$ 
& 4$\rm {}^b$& 12$\rm {}^c$& 3& 3$\rm {}^d$& 4$\rm
{}^d$& 1$\rm {}^d$& 2$\rm {}^d$& 2$\rm {}^d$& 0& 0.263\\ 
$P\mu$ & 5$\rm {}^b$& 10$\rm {}^c$& 3& 4& 3$\rm {}^d$& 2& 2$\rm
{}^d$& 0& 1& 0.771\\ 
tube-tri & 0& 6$\rm {}^a$& 6& 0& 0& 6$\rm {}^b$&
1& 0& 2 $\rm {}^d$
& 1.163\\ 
tube-$\sigma$ $\rm {}^a$& 5$\rm {}^b$& 7$\rm {}^c$& 3& 5$\rm {}^b$&
2$\rm {}^b$& 2& 0& 3$\rm {}^b$& 0& 0.936\\
\end{tabular}
\]
\label{t-geom}
\end{table}

Table~\ref{t-geom} summarizes geometric
data on the coordination environments of these $Z6$
structures. 
~\footnote {
In the table, footnote 
$a$ means nearest ($\sim R$) neighbours are not at identical distances;
$b$ means $N_2$ has a tolerance of $0.14 R$ in distance;  
$c$ means neighbors at various distances 
between $\sqrt 2 R$ and $\sqrt 3 R$ are binned in $N_3$;
$d$ means $\theta_a$ has a tolerance of $6^\circ$ in angle, 
and the ``$180^\circ$'' bin in tube-tri is actually at $169^\circ$. }
Here $N_2$ and $N_3$ are the number of neighbours at
at distances $\sim \sqrt{2} R$ and
$\sim \sqrt{3} R$, respectively. 
The other columns directly determine our potential (\ref{eq-Z}):
the number $N(\theta_a)$ in each coordination shell of 
``two-neighbour'' angles $\theta_a$, 
and the asymmetry $\xi$ of the coordination shell
defined by  (\ref{eq-asym}).

\section{LDA Results}
We performed an {\em ab initio} total energy calculation for each structure
in our database, in the local density approximation
with extended norm and hardness conserving pseudo-potentials~\cite{teter}.
We used all
 planewaves with kinetic energy up to 54.5 Ry.  The Brillouin zone is
sampled with a k-point density of at least $(16.3\AA^{-1})^3$.
Band structure energies are converged to within $10^{-6}$ Ry in each
calculation. The convergence with respect to k-point density shows a
precision of 0.03 eV for energy comparison among different structures.

\begin{figure}
\epsfxsize=5.40in {\epsfbox{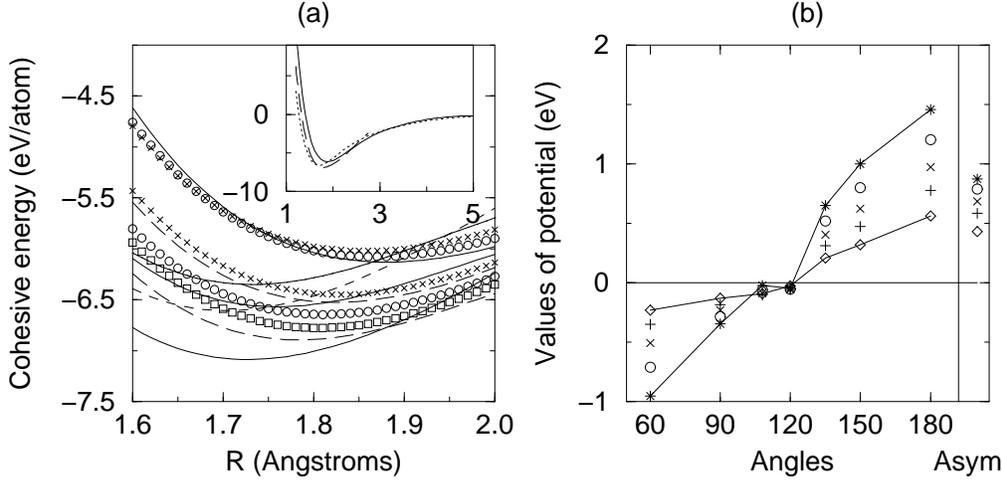}}
\caption {(a). Cohesive energy $E_i(R)$ for $Z6$ structures.  
 In the inset, the
 three representative structures are SC (line), tri (dotted), and
 SC-ico (long-dash). The region showing the energy wells is magnified
 in the main figure.  The structures are, in order of lowest to
 highest energy at $R = 1.8\AA$, ideal-$\alpha_{12}$ (line), SC-ico
 (long-dash), SC-co ($\Box$), tube-$\sigma$ ($\circ$), T-lattice
 (line), P-$\mu$ (long-dash), tri (short-dash), P-$\sigma$ ($\times$),
 tube-tri (line), SC (line), i-dodec ($\circ$), and Kagome
 ($\times$).
(b). Angular and asymmetry potentials.
The fitted $G(R,\theta)$ is plotted with symbols
$\{ *, \circ, \times, +, \diamond \}$, corresponding to $R= \{ 1.6,
1.7, 1.8, 1.9, 2.0 \} \AA$; the first and last sets are
connected by solid and dashed lines, respectively.
The coefficient $c(R)$ (see eqs.~(\ref{eq-Z}) and (\ref{eq-asym})
is shown by the same symbols, on the same scale, in the ``Asym'' column.}
\label{f-Er}    
\label{f-pot}   
\end{figure}

We varied the lattice constant of each structure (a uniform
scale factor, without any relaxation of positions). 
Figure ~\ref{f-Er}(a) collects
 the cohesive energy (per atom) of each $Z6$ structure as a function
of the nearest neighbour distance $R$.
The inset shows the single-well-like shape
for $R$ up to $5 \AA$
for three representative structures, SC, tri, and SC-ico;
beyond $R = 3 \AA$, the curves for all structures converge, approaching
the non-bonding limit.
\rem{
For each structure, with unit length defined as $1.631 \AA$, the
shortest bond length found in the $\alpha_{12}$ phase, we sampled 10
energy data points at increment of 2\% dilation within the single-well
region, and then at total dilations of 0.7, 0.8, 0.9, 1.5, 2.0, 3.0
and 5.0.  We present here the spline fitted curve.}
The main figure enlarges the physically relevant range
$R \in [1.6, 2.0] \AA$, 
to highlight the differences among the structures.  
The lowest energy curve shown is the ``ideal-$\alpha_{12}$'' phase,
topologically same as $\alpha_{12}$ but with all nearest neighbour
distances set equal. (This was omitted from our $Z6$ database because
it has two kinds of sites, of which one has $Z=7$.)
The $Z6$ uniform structure of lowest energy is SC-ico, 
another way of connecting $\rm B_{12}$ icosahedra so each atom
has one inter-icosahedral bond.
Slightly higher is SC-co, which can be viewed analogously as
a packing of $\rm B_{12}$ cuboctahedra in place of
icosahedra,  with each atom having two inter-cluster bonds.
\OMIT {A cuboctahedron is similar to an icosahedron
in molecular orbitals theory \cite{bullett}.}
\OMIT {Finally, the
SC, T-lattice and tri energy curves almost fall on top of each other.}

\section{Fitting of an effective potential}
The simplest effective potential would be 
a single-well two-body potential of typical shape,
so the nearest-neighbour radius lies close to
the minimum of the well, while further neighbours contribute at the
tail of the well.  Then structures of the same coordination 
would have similar energies (from the near-neighbor terms) with 
small differences due to the farther neighbors. 
This will not work in general~\cite{bazant} and failed badly
for our energy curves.  
\OMIT {Numerical fittings (for $Z=4,5,6,7$) with
several single-well forms  bore out this finding.}
Basically, one is asking
the tail of $f_Z(r)$ in (\ref{eq-Z}) to account for
both the total energy 
when $R \to \sqrt 2 R$ or $\sqrt 3 R$ -- 
a sizeable fraction of the well depth --
as well as the dependence on $N_2$ and $N_3$ 
(see Table~\ref{t-geom}) 
among structures of the same $Z$ -- which is smaller, 
as seen in Figure~\ref{f-Er}.  
We conclude that the two-body potential 
must be essentially truncated 
after the nearest-neighbor distance. 
\OMIT { \footnote{
``$P\sigma$ has 4 nearest neighbours at 1.03108 R, and tube-$\sigma$ has 2
at 1.0072 R  and 1 at 1.0256 R.  These corrections are accounted for
in the two-body energies'' --WJZ.
Apparently my (\ref{eq-linrel}) is not literally true -- 
we adopt a reference potential from one structure, and then compute
its contributions to the $f(r)$ terms.  I am worried that this means
the choice of reference structure is not strictly arbitrary -- that
one can get different 3-body potentials at the other end.}  }
It follows that the first term in (\ref{eq-Z}) reduces to $Z f(R)$; 
the energy differences among structures 
of the same coordination $Z$
must be attributed to higher-order potentials.  
For each $Z$, we must choose a reference structure
for which the higher-order terms are set to zero
(thereby {\it defining} $Zf_Z(R)$ to be its energy curve);
for $Z6$ we chose the simple cubic (SC)  structure.

\rem {A major problem in determining the functional form of the angular
potential $g(R,R,\theta)$ is that the frequency of a particular angle
$\theta_a$ cannot be varied independently of $\theta_{a'}$ to display
its individual effect on the cohesive energy.  Normally this problem
is avoided by fitting to an assumed form \cite{bazant}, but for Boron,
where angular behavior is still being explored, we have no analytic
guidance as to the form of this potential.  However, the small number
of distinct angles $\theta_a$ appearing among the structures in
Table~\ref{t-geom} helps to resolve this. }

Thus we are fitting the $E_m(R)$ from Figure~\ref{f-Er}(a)
to the linear equations (\ref{eq-linrel}), in which 
Table~\ref{t-geom} (minus columns $N_2$ and $N_3$) constitutes
the $11\times 8$ matrix of coefficients.
Eqs.~(\ref{eq-linrel}) are overdetermined (by
three degrees of freedom); hence for each $R$, 
we can solve for $G_6(R,\theta_a)$ and $c_6(R)$
in a least-squares sense, without assuming any functional forms.  
The resulting fit (Figure~\ref{f-pot}) shows
the angular potential $G_6(R,\theta)$ increases monotonically with $\theta$.
Both $c_6(R)$ and $G_6(R,\theta)$ show rapid decay as a function of $R$.
Qualitatively similar $(R,\theta)$ dependences 
are found for $Z=4,5,$ and $7$ structures~\cite{PRB}, 
except that for $Z=4$ the sign of $G_4$ is reversed.
The monotonicity of the fitted $G_Z(R,\theta)$, for all four values of $Z$, 
argues for the physical validity of the result:
a spurious fit should produce random fluctuations 
as a function of $\theta_a$, since 
the value of $G_Z$ at each bin is an independent parameter in the fit. 

\section{Discussion}
The first check on our potentials is that, 
among the 11 structures in our $Z6$ database, 
the total energies are in the right order
(apart from exchange of certain close ones), with 
an average error of 0.14 eV/atom (for $R$ in the bonding range).  
Furthermore the ``inverted-umbrella'' coordination shell of
real Boron -- which was {\it not} used in our database --
was correctly found lower
(by $\sim 1$eV) than any of the $Z6$ shells that
{\it were} used.
We also performed a Monte Carlo search of random $Z6$ environments, 
at the valid $R=1.7\AA$ and found the inverted-umbrella was
the second-lowest in energy. 
\footnote{The {\it uninverted} umbrella was lowest in energy, but there 
exists no extended structure in which all atoms have this 
coordination shell.}
Finally, our $G_6$ disfavouring $\theta =$ \deg{180}
(see Fig.~\ref{f-pot}) correctly predicts that
Boron in triangular sheets buckles, as found by {\it ab-initio}
calculations~\cite{boustani}.

Empirical tight-binding calculations are the successful
intermediate between a fully ab-initio and our
classical-potential approach~\cite{lenosky}.
It would be worthwhile to 
fit our database in this fashion, especially
if longer-range, metallic interactions are found within the clusters, 
tubes, and sheets.
Tight-binding (atomic orbital) total energies can also be 
expanded in moments related to circuits of 
three or four atoms.
Structure selection and bond angles was studied in this
way by Lee {\it et al}~\cite{SL}
for selected packings of $\rm B_{12}$ icosahedra.
It would be interesting to 
analytically relate the above-mentioned circuits 
to the form of classical potentials found by us.

Can our potentials, limited to fixed $Z$ and $R$, 
be extended to arbitrary Boron structures? 
First note that our results (Fig.~\ref{f-pot})
are well fit by a separable form $G_Z(R,R,\theta) = g_Z (R)^2 A_Z(\theta)$.
Provided the results are physical reasonable, it 
is straightforward to interpolate
$g_Z$ and $A_Z$ with respect to $R$ and $\theta$, respectively, 
and (less easily) with respect to $Z$.
Then one just needs to replace the $Z$ definition used
in this paper by
one of the well-known formulas 
that depends smoothly on the coordinates, 
e.g. $(\sum _j r_{ij}^{-5} )^2/(\sum _j r_{ij}^{-10})$. 
So finally we would set
$G_Z(r_i,r_j,\theta) \equiv g_Z(r_i) g_Z(r_j) A_Z(\theta)$.
in eq. (\ref{eq-gen}), 
well-defined for 
general structures (provided $4 \leq Z\leq 7$ for all atoms);
this would reduce to our present results (\ref{eq-Z}) within
each family of $Z$-coordinated uniform structures. 
\OMIT {However, these further kinds of fitting have not yet been carried out.
Nor will this extended potential necessarily be {\it correct}
for structures containing differing bond lengths or coordination
numbers, until these are incorporated into the database.}

An attractive application of classical potentials 
would be in the liquid phase.
Here {\it ab initio} molecular dynamics studies 
(on a 48 atom system)
found detailed results for bond-angle distributions and
other correlation functions~\cite{vast95}, 
in good agreement with experiment~\cite{Krish98}.
Those authors also note~\cite{vast95}
that $Z=6$ but the icosahedra 
are broken up (the inverted-umbrella coordination shell is no
longer found), contrary to earlier thought.
Potentials would offer a chance to model 
the structure of amorphous $a$-B, in which few
details are known but (from radial distribution functions)
the $\rm B_{12}$ icosahedron is believed to be the
motif~\cite{Ko97}. 
It would be interesting to explore more ordered
(e.g. micro-quasicrystalline) or less ordered
(like the liquid) variants of this picture. 
Finally, if our method were extended to
extract a classical potential from Si and B/Si structures, 
this could be applied to the icosahedral~\cite{Yam97} or 
cuboctahedral~\cite{Oka97} $\rm B_{12}$ clusters 
which are believed to precipitate in B-doped Si.

To conclude, we have presented a novel approach to generating a
database for potential fitting which includes so many structures
that  the angular potential may be fitted rather than assumed. 
We obtain a reasonable description using potential terms that are
{\it local} to the coordination shell of each atom, and
which could be extended to general Boron structures. 
Modeling of the speculative or poorly-known 
quasicrystal, amorphous, unsolved crystal, liquid,
and Si inclusion forms of Boron would profit from such a potential.

\stars
We thank M. Teter, D.~Allen,  and J.~Charlesworth for providing 
code and support in the LDA calculation, A. Quandt and M. Sadd for  comments,
K. Shirai and  P. Kroll 
for discussions.
This work was supported by DOE grant DE-FG02-89ER45405, 
and used computer facilities of the Cornell Center
for Materials Research supported by NSF grant DMR-9632275.


\vskip-12pt


\begin{thebibliography}{10}


\bibitem{donohue}
J. Donohue, {\it The structures of the elements}
(Wiley, New York, 1974). 

\OMIT {
\bibitem{hoard}
J. L. Hoard and R. E. Hughes, in {\it The Chemistry of Boron and its
Compounds}, edited by E. L. Muetterties (John Wiley \& Sons, New York,
1967), p. 40.  }

\OMIT {
\bibitem{SciAm}
D. B. Sullenger and C. H. L. Kennard, Sci. Am. p. 66, July 1966.}

\bibitem{slack}
G. A. Slack {\em et al},  J. Solid State Chem. {\bf 76}, 52 (1988).
This crystal was previously modeled in a 1/3 smaller cell with 105 atoms.

\OMIT {\bibitem{bullett}
D. W. Bullett, p. 249 in {\it Boron-rich Solids}, edited by D. Emin, et
al. (American Institute of Physics, New York, 1986).}


\bibitem{kleinman92}
I.~Morrison, D.~M.~Bylander, and L.~Kleinman, 
Phys. Rev. B {\bf 45}, 10872 (1992);
S. Lee {\em et al}, Phys. Rev. B {\bf 45}, 3248 (1992).

\bibitem{longuett}
H. C.  Longuet-Higgins and M. de V. Roberts, 
Proc. Roy. Soc. London, Ser. {\bf A 230}, 110 (1955).

\OMIT {\bibitem{wells}
A. F. Wells, {\it Three-dimensional nets and polyhedra}
(John Wiley \& Sons, New York, 1977).}

\bibitem{Beckel}
C.~L.~Beckel
{\it et al}, 
Phys. Rev. B 44, 2535 (1991). 

\bibitem{shirai}
 K. Shirai, Phys. Rev. B {\bf 55} 12235 (1997);
 J. Solid State Chem. {\bf 133}, 215 (1997).

\bibitem {Fu99}
M.~Fujimori, {\it et al}, 
Phys. Rev. Lett. 82, 4452 (1999).

\bibitem {La99}
R.~Lazzari, {\it et al},
Phys.~Rev.~Lett. 83, 3230 (1999).

\bibitem{Ta95}
M.~Takeda {\it et al}, 
p. 739 in {\it Proc. 5th Int'l Conf. on Quasicrystals,}
eds. C. Janot and R.~Mosseri (World Scientific, 1995).


\bibitem{wey95}
D.~Weygand and J.-L. Verger-Gaugry,
C. R. Acad. Sci. II 320, 253 (1995)

\bibitem{kimura}
M. Takeda {\em et al}, \rem{K. Kimura, A. Hori, H. Yamashita, and H. Ino}
Phys. Rev. B {\bf 48} 13159 (1993).


\bibitem{boustani}
I.~Boustani,  A. Quandt, and P. Kramer,
Europhys. Lett., {\bf 36}, 583 (1996);
I.~Boustani and A. Quandt, 
Europhys. Lett., {\bf 39}, 527 (1997).


\bibitem{mailhiot}
C.~Mailhiot {\it et al},
Phys. Rev. B {\bf 42}, 9033 (1990).


\bibitem {vast95}
N. Vast, S. Bernard, and G. Zerah, 
Phys. Rev. B52, 4123 (1995). 

\bibitem{SL}
S. Lee, R. Rousseau, and C. Wells,
Phys. Rev. B {\bf 46}, 12121 (1992).

\bibitem{stillinger}
F.~H. Stillinger and T.~A. Weber,
Phys. Rev. B {\bf 31}, 5262 (1985).


\bibitem{tersoff}
J. Tersoff, Phys. Rev. Lett. {\bf 56}, 632 (1986); 
Phys. Rev. B {\bf 38}, 9902 (1988).


\bibitem{chelikowsky}
J.~R. Chelikowsky, K.~M. Glassford, and J.~C. Phillips,
Phys. Rev. B {\bf 44}, 1538 (1991);
Phys. Rev. Lett. {\bf 62}, 292 (1989).


\bibitem{bazant}
M. Z. Bazant and E. Kaxiras, Phys. Rev. Lett. {\bf 77}, 4370 (1996).

\OMIT {\bibitem{okeefe}
M. O'Keefe and B.G. Hyde, {\it Crystal Structures} (Mineralogical
Society of America, Washington, D.C. 1996),  p. 240, 276. }



\bibitem{PRB}
W.-J. Zhu and C. L. Henley, unpublished.

\bibitem{teter}
M. C. Payne {\it et al},  Rev. Mod. Phys. {\bf 64}, 1045 (1992);
M. Teter, Phys. Rev. B {\bf 48}, 5031 (1993).


\bibitem {lenosky}
T. J. Lenosky {\it et al}, 
Phys. Rev. B 55, 1528 (1997). 


\bibitem{Krish98}
S.~Krishnan, 
{\it et al}, 
Phys. Rev. Lett. 81, 586 (1998).


\bibitem{Ko97}
M. Kobayashi,   I. Higashi, and M. Takami M
J. Solid State Chem. 133, 211 (1997).


\bibitem {Yam97}
J.~Yamauchi, N.~Aoki, and I.~Mizushima 
Phys. Rev. B 55: 10245 (1997).



\bibitem {Oka97} 
M.~Okamoto, K.~Hashimoto, and K.~Takayanagi,
Appl. Phys. Lett. 70, 978 (1997).






\end{thebibliography}
\end{document}